\documentclass[amsmath, amsfonts, twocolumn, prb]{revtex4}
\usepackage{graphicx}
\usepackage{epsfig}
\usepackage{bm}
\usepackage{dcolumn}
\usepackage{amsmath}
\usepackage{amssymb}
\usepackage{color}
\usepackage[varg]{txfonts}

\def\nn{\nonumber}
\newcommand{\dett}{\mathop{\rm Det}\nolimits}
\newcommand{\res}{\mathop{\rm Res}\nolimits}
\newcommand{\rme}{\mathrm{e}}
\newcommand{\rmh}{\mathrm{h}}
\newcommand{\trasp}{\mathsf{T}}

\begin{document}

\title{Weyl nodes in Andreev spectra of multiterminal Josephson junctions: \\  Chern numbers, conductances and supercurrents}

\author{Hong-Yi Xie}
\affiliation{Department of Physics, University of Wisconsin-Madison, Madison, Wisconsin 53706, USA}

\author{Maxim G. Vavilov}
\affiliation{Department of Physics, University of Wisconsin-Madison, Madison, Wisconsin 53706, USA}

\author{Alex Levchenko}
\affiliation{Department of Physics, University of Wisconsin-Madison, Madison, Wisconsin 53706, USA}

\date{July 24, 2017}

\begin{abstract}
We consider mesoscopic four-terminal Josephson junctions and study emergent topological properties of the Andreev subgap bands. We use symmetry-constrained analysis for Wigner-Dyson classes of scattering matrices to derive band dispersions. When scattering matrix of the normal region connecting superconducting leads is energy-independent, the determinant formula for Andreev spectrum can be reduced to a palindromic equation that admits a complete analytical solution. Band topology manifests with an appearance of the Weyl nodes which serve as monopoles of finite Berry curvature. The corresponding fluxes are quantified by Chern numbers that translate into a quantized nonlocal conductance that we compute explicitly for the time-reversal-symmetric scattering matrix. The topological regime can be also identified by supercurrents as Josephson current-phase relationships exhibit pronounced nonanalytic behavior and discontinuities near Weyl points that can be controllably accessed in experiments.       
\end{abstract}

\maketitle

\section{Introduction}

Dirac [\onlinecite{Dirac}], Weyl [\onlinecite{Weyl}], and Majorana [\onlinecite{Majorana}] fermions of relativistic field equations have recently entered the realm of modern condensed matter physics, see reviews [\onlinecite{Kane,Qi,Vafek,Elliott,Beenakker-RMP,Armitage}]. These theories provide an effective Hamiltonian description for linearly dispersing low-energy quasiparticle excitations near symmetry-enforced or accidental degeneracy points that occur in electronic band structures. This remarkable connection to solid state systems enables testing predictions of fundamental particle physics, such as for example Adler-Bell-Jackiw chiral anomaly  [\onlinecite{Adler,Bell-Jackiw}], by means of transport measurements via chiral magnetic effect or magnetoresistance (for a comprehensive contemporary review on transport consequences of anomalies see Ref. [\onlinecite{Burkov}]).  Moreover, the fact that the strict symmetries of the free space do not necessarily hold in a lattice opens additional perspectives that have no counterparts in the high-energy physics. In particular, the lack of strict Lorentz invariance allows for a tilt of the Weyl cones that could result in a touching between electron and hole pockets forming an open Fermi surface. These systems were termed type-II Weyl semimetals [\onlinecite{Soluyanov}]. Furthermore, crystal symmetries can stabilize other degenerate band touchings leading to unconventional fermions [\onlinecite{Wang,Bradlyn}]. For example, in the presence of time reversal symmetry the three fold degeneracies can occur as a consequence of nonsymmorphic symmetries or even in symmorphic structures provided a combined rotation and mirror symmetries. To leading order, these points are formed by two linearly dispersing bands bisected by a flat band and can be effectively regarded as spin-1 Weyl fermions. These systems are also topologically distinct as they carry Chern numbers equal to $\pm 2$ as compared to $\pm1$ in the usual case of Weyl nodes. 

Mathematical structure of the band theory, with its link to topology via the Berry curvature constructed from the Bloch states, which defines Chern integral invariants, inspired the search for rich topological states in non-electronic systems. The examples include photonic crystals that realize electromagnetic analog of Dirac nodes [\onlinecite{Haldane,Raghu}], acoustic and optomechanical systems [\onlinecite{Peano}], and ultracold atoms in three-dimensional optical lattices [\onlinecite{Dubcek}]. The attention has recently turned to superconducting systems. Superconductivity in Weyl semimetals was discussed theoretically [\onlinecite{Meng,Cho,Bednik}] even prior to the candidate material was proposed, and observed only recently in transition metal dichalcogenides WTe$_2$, MoTe$_2$, PdTe$_2$ [\onlinecite{Qi-NC,Noh}]. In a parallel avenue of developments, a combination of Weyl systems with superconductors in a proximity effect was considered in multiple studies [\onlinecite{Chen-EPL,Uchida,Khanna,Chen-PRB,Kim-PRB,Baireuther,Madsen}]. 

It was further realized that even in the hybrid structures made of mundane superconducting materials the topology intrinsic to Weyl systems can still emerge. This is possible in the geometry of three- and four-terminal Josephson junctions [\onlinecite{Akhmerov,Yokoyama,Riwar,Eriksson,Houzet,HYX}]. Indeed, at each superconductor-normal interface Andreev reflections convert electron-like to hole-like excitations, and vise versa, that leads to a formation of localized subgap states. In a two-terminal case of a short junction, when junction length is much smaller than superconducting coherence length, such states form a one-dimensional band per each conducting mode in a junction which is determined by its respective transmission eigenvalue. The mode dispersion is governed by the dependence of the Andreev level on the superconducting phase difference across the junction that plays the role of effective quasi-momentum. 
Periodicity of Andreev level energy on the superconducting phase modulo $2\pi$ mimics Brillouin zone thus making a connection to band theory more transparent. In a three-terminal junction, such Andreev bound states (ABS) form a two-parametric family of bands, whereas a four-terminal junction is equivalent to a three-dimensional artificial solid. Both three- and four-terminal junctions can realize Weyl singularities in ABS spectra, however, their appearance is directly linked to a symmetry properties of the scattering matrix connecting superconducting terminals. In three-terminal devices, the time reversal symmetry must be broken, so that scattering matrix should belong to the circular unitary ensemble (CUE). While it is established that existence of the Weyl nodes in a crystal is ultimately connected to broken either time reversal or inversion symmetry, the conventional classification of topological semimetallic phases does not directly apply to multiterminal superconducting devices. In a four-terminal structure, Weyl nodes of ABS appear already for the time reversal symmetric scenario of an energy independent scattering matrix belonging to circular orthogonal ensemble (COE), which effectively corresponds to a zero-dimensional case. Another comparative point between crystalline and artificial systems concerns the Nielsen-Ninomiya theorem [\onlinecite{NN}] that dictates that net chirality is zero so that Weyl nodes appear in pairs. However, there may be situations where a pair of Weyl nodes of opposite chirality coexists at a point in the three-dimensional Brillouin zone -- and therefore four bands touch, rather than two. This scenario corresponds to Dirac semimetals that can be also realized in multiterminal Josephson junctions as a consequence of particle-hole symmetry and additional Kramers degeneracy of Andreev levels. An account of spin-orbit effects can lift Kramers degeneracy by coupling spin of the bound states to the superconducting phase difference.   

These intriguing possibilities to create and manipulate various topological states harbored by ABS in Josephson contacts set the stage for our work. In Sec. \ref{Sec-ScatteringMatrix} we carry detailed symmetry analysis of the determinant formula that defines subgap structure of Andreev levels in terms of properties of the scattering matrix. A convenient parametrization of the matrix is also presented. We demonstrate that in a simplifying case of the energy-independent scattering matrix an exact analytical results can be found.  We build on these findings to discuss emergent topology of Andreev bands in Sec. \ref{Sec-ABS} with the emphasis on four-terminal devices. We demonstrate formation of Weyl and Dirac singularities, construct their Berry fluxes and compute corresponding Chern numbers. In Sec. \ref{Sec-Jc-G} we explore nonlocal conductance as a practical measure of topological phases. We also calculate Josephson current-phase relationship (CPR) for various trivial and topological states that may provide an additional way to identify them in experiments. We close in Sec. \ref{Sec-SumDis} with the summary of main results and perspectives for the future research.      

\section{Scattering Matrix Formalism}\label{Sec-ScatteringMatrix}

\subsection{Symmetry analysis}

\begin{figure}
\includegraphics[width=0.35\textwidth]{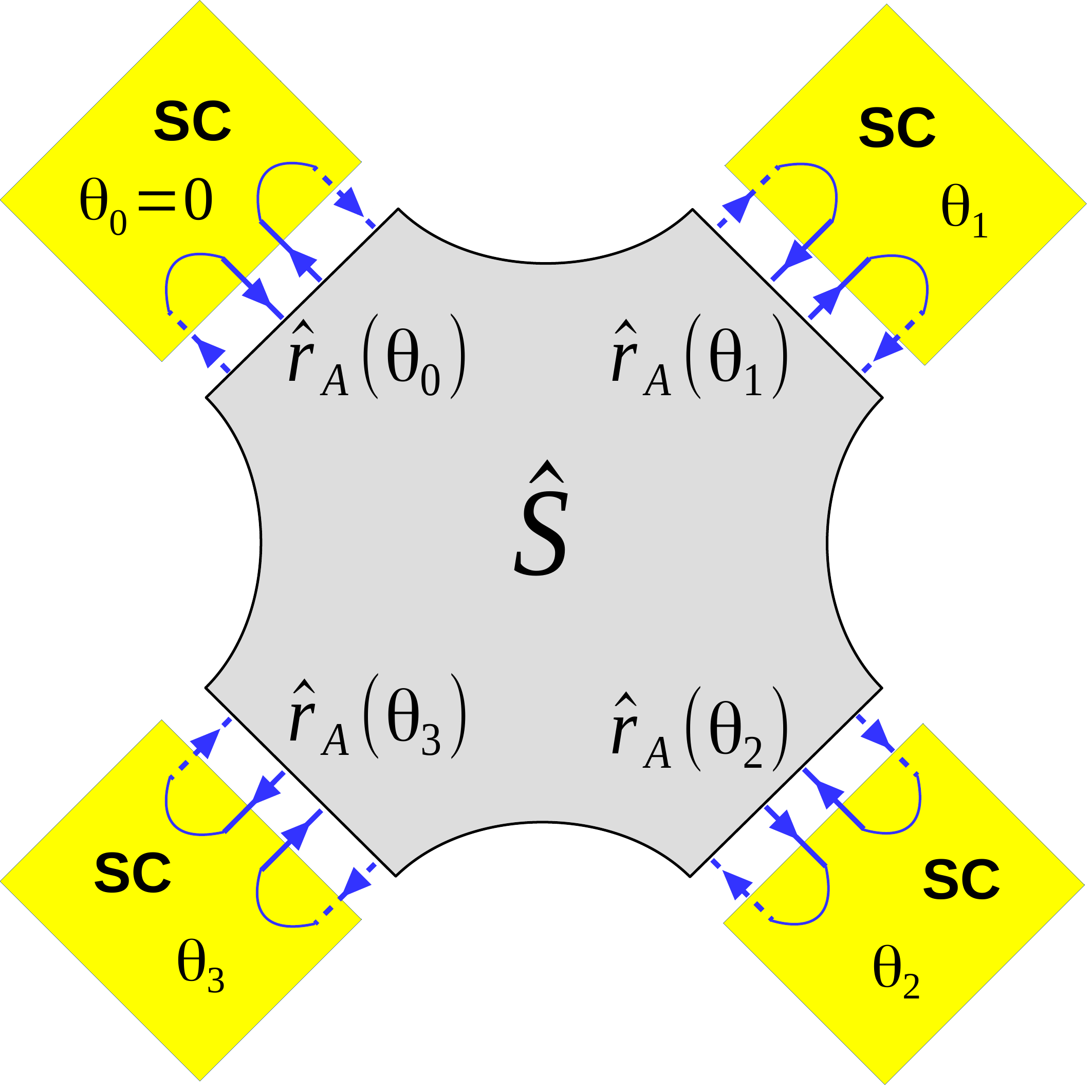} 
\caption{[Color online] Schematic illustration of a four-terminal Josephson junction. The junction is described by a scattering matrix $\hat{s}$. The processes of Andreev reflections taking place at superconducting leads (SC) are represented by corresponding matrices $\hat{r}_A$. We assume the same magnitude of the order parameter in all leads while the phases $\theta_0,\ldots,\theta_3$ can be different.} 
\label{Fig-setup}
\end{figure}

Consider $n$-terminal Josephson junctions (an example of a four-terminal junction is shown in Fig. \ref{Fig-setup}). The coupling between the superconducting leads through the normal region is fully characterized by the electron scattering matrix $\hat{S}(\varepsilon)$, with $\varepsilon$ being the excitation energy. In what follows we assume that all leads have the same superconducting gap $\Delta$ and thus normalize all energies in units of $\Delta$. In general, each lead can be connected to a normal region by a multiple conducting channels that in a given mesoscopic devices are described by a set of random transmission coefficients. To keep the presentation simple, we assume that each junction supports only one channel. Inclusion of multiple channels complicates the analytical structure of the theory, but already our simple model captures the most significant physics.     

When $|\varepsilon|<1$ an electron (e) striking the interface with the superconductor
must be reflected back as a hole (h) via Andreev refection process. The corresponding reflection coefficient picks the phase $\theta$ of the order parameter in that lead. In a subsequent scattering this hole will be converted back at the other interface. The resulting trajectories of electron and hole superpositions form Andreev bound states in the junctions those locations within the energy gap $\varepsilon\in[-1,1]$ depend on all phases of superconducting order parameters. Mathematically, ABS energies are determined by the  
determinant equation [\onlinecite{Beenakker-PRL}]
\begin{equation}\label{ABS-Det-1}
\dett\left[ \mathbb{I}_{4 n}-\hat{R}_\mathrm{A}(\varepsilon,\hat{\theta}) \hat{S}(\varepsilon) \right]=0,  
\end{equation}
where $\hat{R}_A(\varepsilon,\hat{\theta})$ is the scattering matrix describing Andreev reflections and $\hat{\theta}=\{\theta_0,\theta_1,\ldots,\theta_{n-1}\}$ is the diagonal matrix of superconducting phases. In the basics of the $4n$-component Nambu spinor $\psi_\alpha = \left( \psi_{\rme,\alpha,\uparrow}, \psi_{\rme,\alpha,\downarrow},  \psi_{\rmh,\alpha,\uparrow}, \psi_{\rmh,\alpha,\downarrow} \right)^\trasp$ with $\alpha =0,\cdots,n-1$, indicating the leads, the scattering matrix is block-diagonal 
\begin{equation}
\hat{S}(\varepsilon)=\left[\begin{array}{cc} \hat{s}_\rme(\varepsilon) & 0 \\ 0 & \hat{s}_\rmh(\varepsilon)\end{array}\right],
\end{equation}
since electron and hole states in the normal region are not coupled. The intrinsic particle-hole symmetry ($\mathcal{P}$) of the Bogolubov-de Gennes equation imposes an additional constraint  
\begin{equation}\label{PHS}
\mathcal{P}:\quad \hat{S}(\varepsilon)=\hat{\tau}_2\hat{S}^*(-\varepsilon)\hat{\tau}_2, 
\end{equation} 
which implies that $\hat{s}_\rmh(\varepsilon)=\hat{s}^*_\rme(-\varepsilon)$, where $\hat{\tau}_{1,2,3}$ are Pauli matrices operational in the particle-hole $\mathcal{P}$-space. In addition the time-reversal symmetry $(\mathcal{T})$ is represented by
\begin{equation} \label{TRS}
\mathcal{T}: \quad \hat{S}(\varepsilon) = \hat{\sigma}_2 \, \hat{S}^\trasp(\varepsilon) \, \hat{\sigma}_2,
\end{equation}
where $\hat{\sigma}_{1,2,3}$ act in spin space, so that $\hat{s}_{\rme(\rmh)}(\varepsilon) = \hat{\sigma}_2  \hat{s}_{\rme(\rmh)}^\trasp(\varepsilon)  \hat{\sigma}_2$. The Andreev scattering matrix reads
\begin{subequations}
\begin{align}
& \hat{R}_\mathrm{A}(\varepsilon, \hat{\theta}) = e^{- i \arccos{\varepsilon}} \begin{bmatrix} 0 & \hat{r}_{\rme \rmh}(\hat{\theta}) \\ \hat{r}_{\rmh \rme}(\hat{\theta}) & 0 \end{bmatrix}, \\
& \hat{r}_{\rme \rmh}(\hat{\theta}) =  e^{i \hat{\theta}} \otimes i \hat{\sigma}_2, \quad \hat{r}_{\rmh \rme}(\hat{\theta}) = -\hat{r}_{\rme \rmh}(-\hat{\theta}).
\end{align}
\end{subequations}
The particle-hole and time-reversal symmetries are represented as
\begin{subequations}
\begin{align}
& \mathcal{P}: \quad \hat{R}_\mathrm{A}(\varepsilon, \hat{\theta}) = -\hat{\tau}_2 \, \hat{R}_\mathrm{A}^*(-\varepsilon, \hat{\theta}) \, \hat{\tau}_2, \\
& \mathcal{T}: \quad \hat{R}_\mathrm{A}(\varepsilon, \hat{\theta}) = \hat{\sigma}_2 \, \hat{R}_\mathrm{A}^\trasp(\varepsilon, -\hat{\theta}) \, \hat{\sigma}_2.
\end{align}
\end{subequations}
These relations allow to reduce Eq. \eqref{ABS-Det-1} to the form 
\begin{equation}\label{ABS-Det-2}
\mathrm{Det}\left[ \mathbb{I}_{2 n} - \gamma(\varepsilon) \, \hat{r}_{\mathrm{A}}(\hat{\theta}) \, \hat{s}^*(-\varepsilon) \,\hat{r}_{\mathrm{A}}(-\hat{\theta}) \, \hat{s}(\varepsilon)
\right]=0, 
\end{equation}
where $\gamma(\varepsilon)=e^{- 2i \arccos{\varepsilon}}$, and we have defined $\hat{s}(\varepsilon) \equiv \hat{s}_{\rme}(\varepsilon)$ and $\hat{r}_{\mathrm{A}}(\hat{\theta}) \equiv  e^{i \hat{\theta}} \otimes \hat{\sigma}_2$. The left-hand-side determinant can be expanded into a degree-$2n$ characteristic polynomial of $\gamma(\varepsilon)$,
\begin{equation}\label{def-poly} 
P_{2 n} (\gamma; \hat{\theta},\varepsilon) \equiv \sum_{i=0}^{2 n} a_i (\hat{\theta},\varepsilon) \, \gamma^i(\varepsilon), \quad a_0 = 1,
\end{equation}
with the constraint $a_i^*(\hat{\theta},-\varepsilon) = a_i(\hat{\theta},\varepsilon)$ because $P_{2 n}^* (\gamma; \hat{\theta},-\varepsilon) = P_{2 n} (\gamma; \hat{\theta},\varepsilon)$. The coefficients $\{a_i(\hat{\theta},\varepsilon)\}_{i=0}^{2n}$ encode all the system parameters, such as the energy $\varepsilon$, the phases of the superconducting terminals $\hat{\theta}$, and the normal-region scattering matrix properties. When the scattering matrix $\hat{s}(\varepsilon)$ is energy independent, the coefficients $\{a_i( \hat{\theta},\varepsilon)\}_{i=0}^{2n}$ become real $a_i(\hat{\theta}) = a_i^*(\hat{\theta})$ with $0 \le i \le n$. 
One can readily prove that the polynomial (\ref{def-poly}) is in fact palindromic,
\begin{equation} \label{CUE-2n}
\gamma^{2 n} P_{2n}(1/\gamma;\hat{\theta}) =  P_{2n}(\gamma;\hat{\theta}),
\end{equation}
so that the coefficients satisfy $a_i = a_{2n-i}$ with $0 \le i < n$. The reciprocal relation (\ref{CUE-2n}) generally applies to the systems lacking of time-reversal (\ref{TRS}) and spin-rotation symmetry (in the presence of both magnetic field and spin-orbit interactions), i.e., $\hat{s} \in \mathrm{CUE}(2n)$, and the roots appear in $n$ complex conjugate pairs $\{\gamma_j, \gamma_j^\ast \}_{j=1}^n$. In the presence of time-reversal symmetry (\ref{TRS}) yet without spin-rotation symmetry of symplectic class, one has $\hat{s} \in \mathrm{CSE}(n)$ and the Andreev levels are doubly degenerate due to Kramers theorem. 

In this work we focus on spin-degenerate as well as time-reversal symmetric and energy independent scattering matrices $\hat{s}$ so that the determinant (\ref{ABS-Det-2}) simplifies further to
\begin{equation} \label{ABS-Det-3}
\mathrm{Det}\left[ \mathbb{I}_{n} - \gamma(\varepsilon) \, e^{i \hat{\theta}} \, \hat{s}^\ast \, e^{-i \hat{\theta}} \, \hat{s}\right]=0,  
\end{equation}       
and the characteristic polynomial satisfies $\gamma^n P_n(\gamma^{-1}) = (-1)^n P_n(\gamma)$, which is palindromic for $n \in \mathrm{even}$ and antipalindromic for $n \in \mathrm{odd}$. According to the fundamental theorem for (anti) palindromic polynomials, we draw important properties. (i) For $n \in \mathrm{odd}$, $P_n(\gamma)$ can be factorized into a product of the linear palindromic polynomial $\gamma-1$ and a palindromic polynomial (for $n=3$ see Ref.~[\onlinecite{HYX}]). (ii) For $n \in \mathrm{even}$, there exists a degree-$n/2$ polynomial of the new variable $z = \gamma+1/\gamma$, $Q_{n/2}(z) \equiv \gamma^{-n/2} P_n(\gamma)$. We will make use of these properties in the next section when solving Eq. \eqref{ABS-Det-3} for $n=4$ for a device illustrated in Fig. \ref{Fig-setup}.
  
\subsection{Four-terminal junction}

To make further progress with analytical calculations we need a particular parametrization for the scattering matrices. Of course there is no unique way to parametrize them, as the product of two such matrices being again unitary. In most cases the specific choice of parametrization depends on convenience in which the relevant physical quantities of interest are expressed in simplest form possible. Fortunately, there exists a recursive algorithm allowing the parametrization of matrices of dimension $n$ through the parametrization of matrices of dimension $n -2$, the parametrization of $n-1$ dimensional matrices being directly obtained from it without other computations [\onlinecite{Dita}]. The case of $n=4$ relevant to our study was worked out in detail. It is given with a minimum number of real parameters out of which a part take values in the positive unit cube, the others being arbitrary phases. Specifically, for the time reversal symmetric matrix $\hat{s}$ is determined by ten real parameters, and we adopt the following structure [\onlinecite{Dita}]
\begin{align} \label{S}
& s_{11} = a \, e^{i \varphi_{11}}, \, s_{12}  = b \sqrt{1-a^2}  \, e^{i \varphi_{12}}, \nn \\
& s_{13}  = c \sqrt{(1-a^2)(1-b^2)}  \, e^{i \varphi_{13}}, \nn \\
& s_{14}  = \sqrt{(1-a^2) (1-b^2) (1-c^2)}\,  e^{i \varphi_{14} }, \nn \\
& s_{22}  =  p \, (1-b^2) \, e^{i\varphi_{23}} \left[  c^2 d \, e^{i (\varphi_{22}-\varphi_{23})} -b^2 d (1-c^2)  e^{i (\varphi_{23}-\varphi_{22}) } \right. \nn \\
&  \left. + 2 b c \sqrt{(1-c^2)(1-d^2)}\,\right]- a b^2 \, e^{i (2 \varphi_{12}-\varphi_{11})}, \nn \\ 
& s_{23} = -a b c \sqrt{1-b^2} e^{i (\varphi_{12} + \varphi_{13} - \varphi_{11})} - p \sqrt{1-b^2} \, e^{i (\varphi_{13} - \varphi_{12} + \varphi_{23})} \nn \\
&\, \times \Big\{  b c d  \left[ e^{i (\varphi_{22}-\varphi_{23})} + (1-b^2) (1-c^2) e^{i (\varphi_{23}-\varphi_{22}) } \right] \nn \\
&   + \left[ b^2-(1-b^2) c^2 \right] \sqrt{(1-c^2) (1-d^2)}\,\Big\}, \nn \\ 
& s_{24} = -ab \sqrt{(1-b^2) (1-c^2)}\,  e^{i (\varphi_{12} + \varphi_{14} - \varphi_{11})} \nn \\
 &  -\sqrt{1-b^2} \, e^{i (\varphi_{14} + \varphi_{23} - \varphi_{12})} \left[ c \sqrt{1-d^2}-b d \sqrt{1-c^2} e^{i (\varphi_{23}-\varphi_{22})} \right], \nn \\
& s_{33} =  -a c^2 (1-b^2) \,  e^{i (2 \varphi_{13} - \varphi_{11})}  -p \, e^{i (2 \varphi_{13} + \varphi_{23} - 2\varphi_{12})} \nn \\
&\, \times \left[ -b^2 d \, e^{i (\varphi_{22}-\varphi_{23})} + c^2 d (1-b^2)^2 (1-c^2) \, e^{i (\varphi_{23}-\varphi_{22})} \right. \nn \\
 &  \left. + 2 bc(1-b^2)\sqrt{(1-c^2) (1-d^2)}\, \right], \nn \\ 
& s_{34} = -a c (1-b^2) \sqrt{1-c^2} \, e^{i (\varphi_{13} + \varphi_{14} -\varphi_{11})} \nn \\
& + e^{i (\varphi_{13} + \varphi_{14} + \varphi_{23}-2 \varphi_{12})} \left[b \sqrt{1-d^2} +c d (1-b^2) \sqrt{1-c^2} \, e^{i (\varphi_{23}-\varphi_{22}) } \right], \nn \\
& s_{44} = -a (1-b^2) (1-c^2) e^{i (2 \varphi_{14} - \varphi_{11})} - \frac{d}{p} e^{i (2 \varphi_{14} + 2 \varphi_{23} - 2 \varphi_{12} - \varphi_{22})},
\end{align}
and $s_{ji} = s_{ij}$ for $1 \le i < j \le 4$, where $p =( b^2 + c^2 - b^2 c^2)^{-1}$,  $a,b,c,d \in [0,1]$, and $\varphi_{11,12,13,14,22,23} \in [0, 2\pi)$. Next we analyze Eq. \eqref{ABS-Det-3} in terms of Eq. \eqref{S}. 

\section{Andreev bands and topological characteristics}\label{Sec-ABS}

\subsection{Andreev spectra}\label{Sec-Chern}

In this section we solve Eq. \eqref{ABS-Det-3} for energies of Andreev states as a function of superconducting phases and parameters of the scattering matrix. We construct their respective Bloch states and define Berry curvatures to characterize Andreev band topologies in terms of Chern numbers. We find both trivial and topological states and corresponding quantum phase transitions that can be tuned by adjusting phases in the leads. Closing of the Andreev gap is found to occur either at a single or a pair of points in a parameter space of phases. 

For $n=4$ the palindromic equation \eqref{ABS-Det-3} reads $\gamma^4 - A \gamma^3 + B \gamma^2 - A \gamma + 1 =0$ that determines four Andreev bands
\begin{equation}\label{ABS}
\varepsilon (\boldsymbol{\theta}) = \pm \sqrt{\frac{A(\boldsymbol{\theta}) + 4 \pm \sqrt{A^2(\boldsymbol{\theta}) - 4 B(\boldsymbol{\theta}) + 8}}{8}},
\end{equation}
where $\boldsymbol{\theta} \equiv (\theta_1,\theta_2,\theta_3)$, whereas fourth phase $\theta_0=0$ can be set to zero owing to global gauge invariance. The $A$- and $B$-functions take the form
\begin{subequations} \label{A-B}
\begin{align}
&\,  A = A_0 + 2 \sum_{i=1}^{3} A_i \cos{\theta_i} + 2  \sum_{i<j} A_{ij} \cos(\theta_i - \theta_j), \label{A} \\  
&\,  B = B_0 + 2 \sum_{i=1}^{3} B_i \cos{\theta_i} + 2 \sum_{i<j} B_{ij} \cos(\theta_i - \theta_j) \nn \\ 
                          &\, \quad \quad + 2 \sum_{i j k \in P_{123}} B_{ijk} \cos(\theta_i + \theta_j - \theta_k), \label{B}
\end{align}
\end{subequations}
with permutations $P_{123} =\{123, 312, 231 \}$. In terms of the matrix elements of $\hat{s}$ from Eq. \eqref{S} the coefficients in Eqs.~(\ref{A}) and (\ref{B}) read
\begin{equation}
A_0 = \sum_{j=1}^4 \left| s_{jj} \right|^2, \quad A_i =  \left| s_{1,i+1} \right|^2, \quad A_{ij} =  \left| s_{i+1,j+1} \right|^2,
\end{equation}
and 
\begin{equation}
\begin{split}
& B_0 = \sum_{i<j} \left| s_{ii} s_{jj} -s_{ij}^2 \right|^2, \quad B_{123} = \left| s_{13} s_{24} - s_{12} s_{34} \right|^2  \\
& B_1 =  \left| s_{13} s_{23} - s_{12} s_{33} \right|^2 + \left| s_{14} s_{24} - s_{12} s_{44} \right|^2 , \\
& B_2 =  \left| s_{12} s_{23} - s_{13} s_{22} \right|^2 + \left| s_{14} s_{34} - s_{13} s_{44} \right|^2 , \\
& B_3 =  \left| s_{12} s_{24} - s_{14} s_{22} \right|^2 + \left| s_{13} s_{34} - s_{14} s_{33} \right|^2 , \\
& B_{12} =  \left| s_{12} s_{13} - s_{11} s_{23} \right|^2 + \left| s_{24} s_{34} - s_{23} s_{44} \right|^2, \\
& B_{13} = \left| s_{12} s_{14} - s_{11} s_{24} \right|^2 + \left| s_{24} s_{33} - s_{23} s_{34} \right|^2 , \\
& B_{23} =  \left| s_{13} s_{14} - s_{11} s_{34} \right|^2 + \left| s_{23} s_{24} - s_{22} s_{34} \right|^2, \\
& B_{312} = \left| s_{14} s_{23} - s_{12} s_{34} \right|^2,  \quad B_{231} = \left| s_{14} s_{23} - s_{13} s_{24} \right|^2. 
\end{split}
\end{equation} 
An inspection of these expressions reveal that despite the fact that we need six independent phases to parametrize scattering matrix only two effective angle variables $\varphi_1 \equiv \varphi_{22} - \varphi_{23}$, $\varphi_2 \equiv  \varphi_{11}+\varphi_{22} - 2 \varphi_{12}$ affect the Andreev spectrum in Eq. \eqref{ABS}.

\begin{figure}
\includegraphics[width=0.23\textwidth]{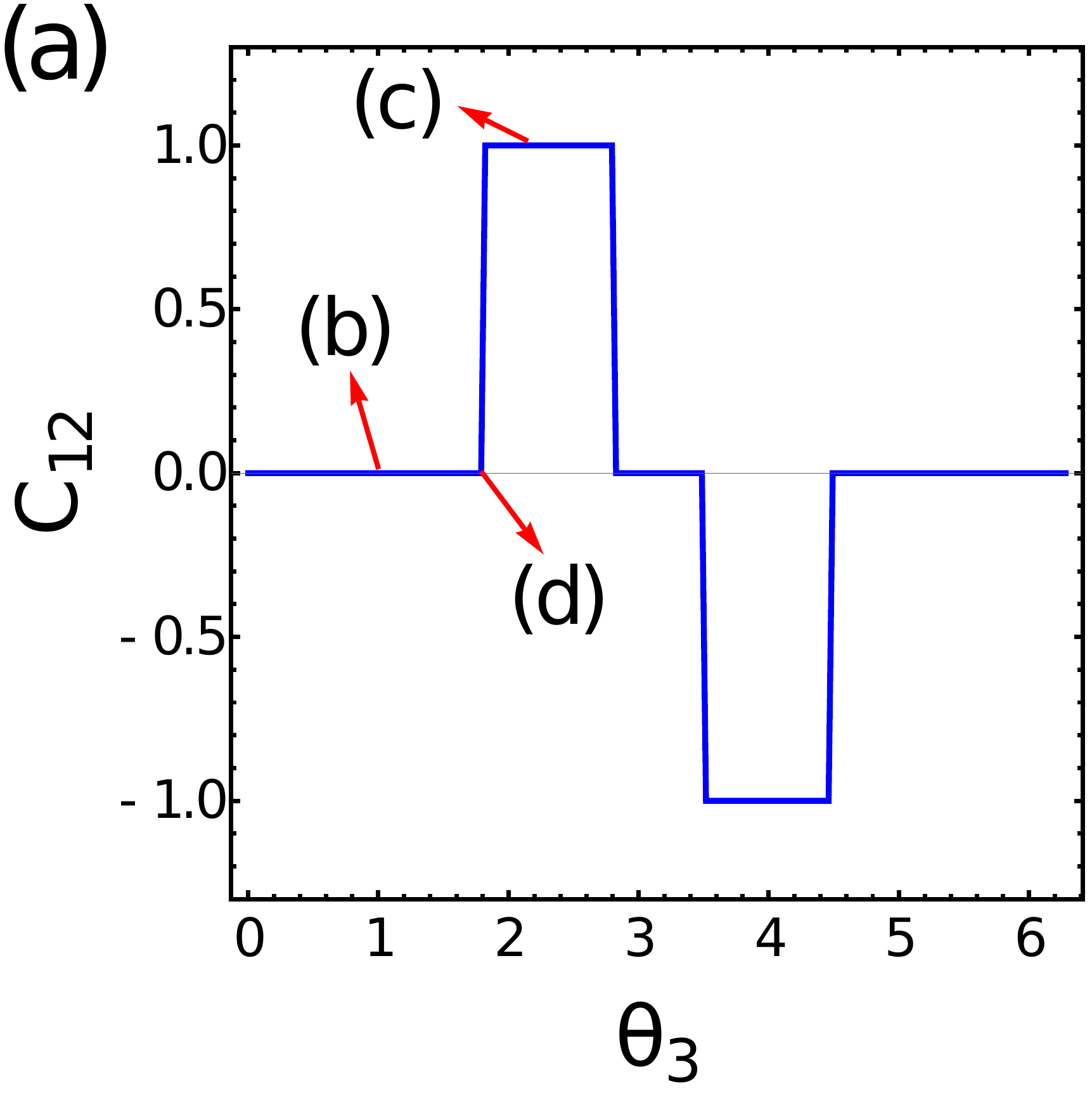} 
\includegraphics[width=0.23\textwidth]{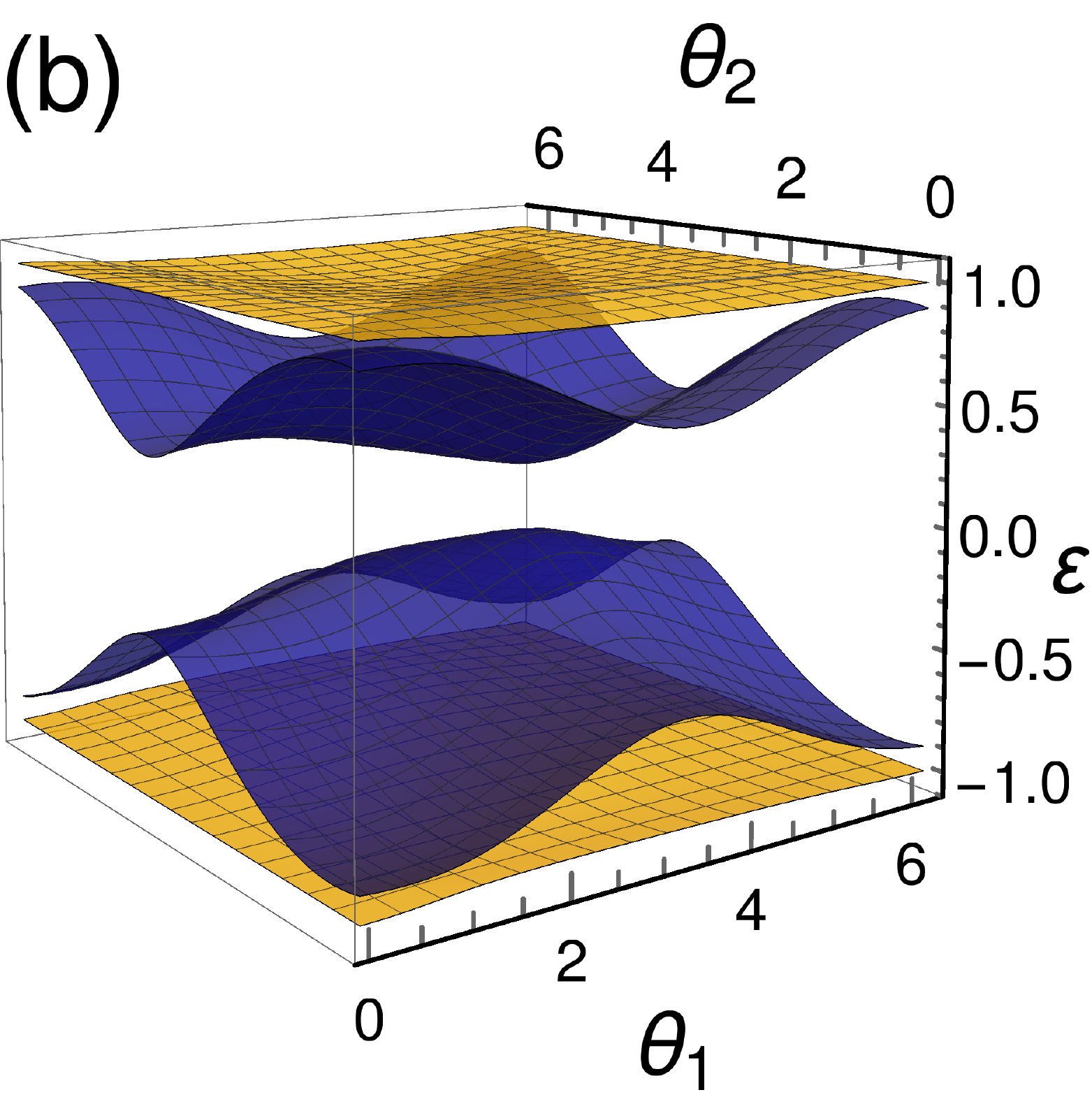}    \\
\includegraphics[width=0.23\textwidth]{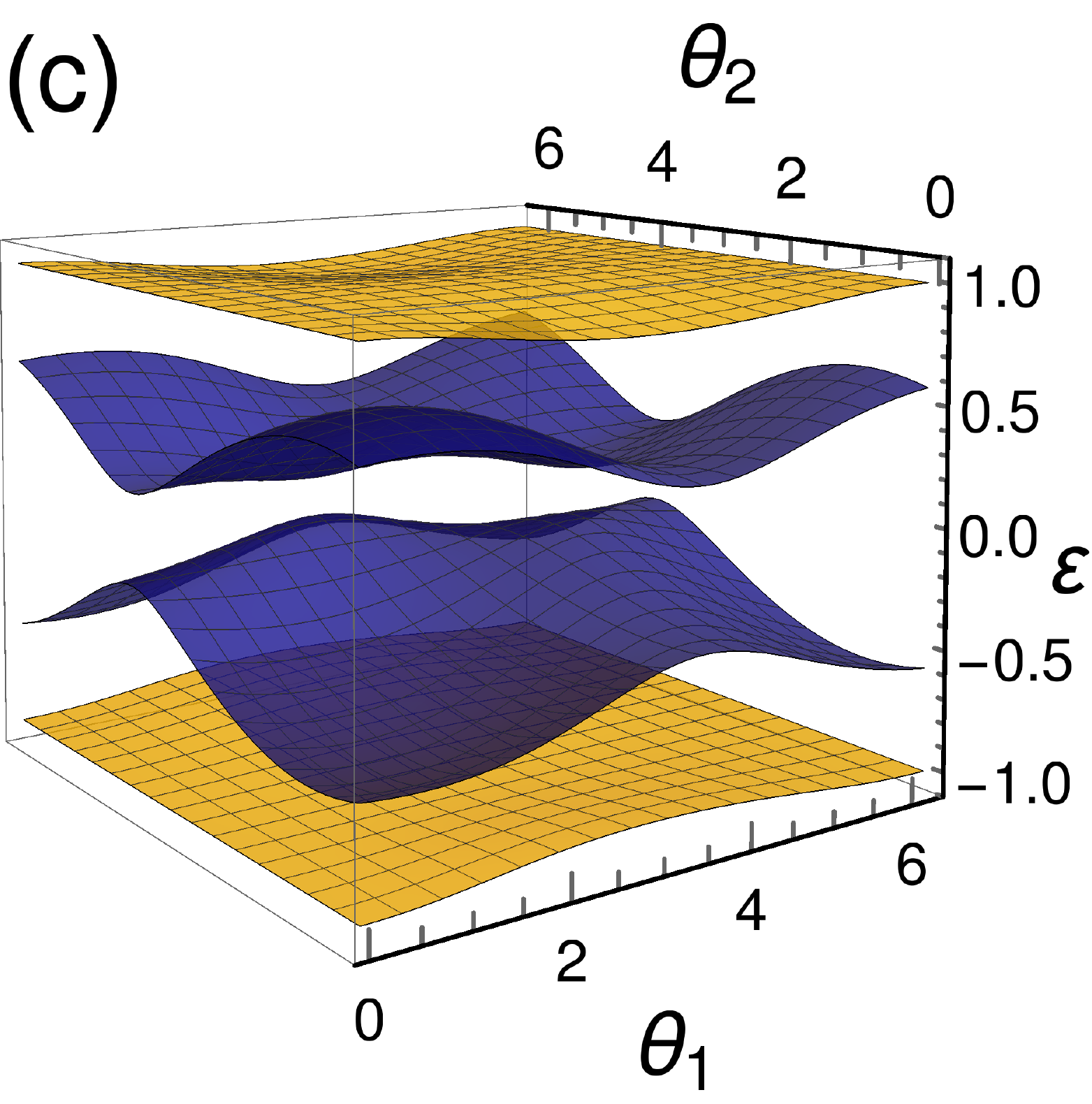}
\includegraphics[width=0.23\textwidth]{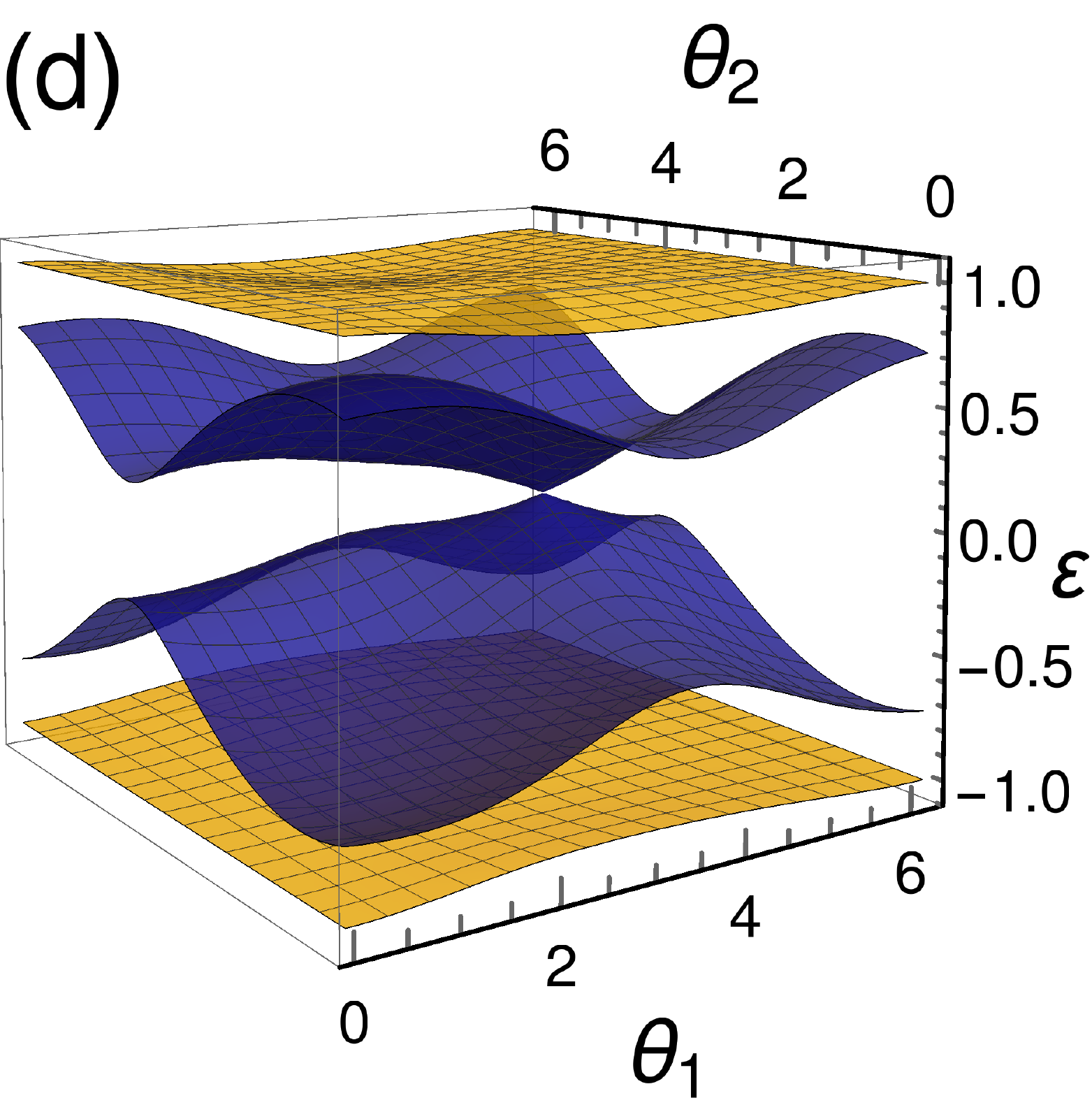}  \\
\includegraphics[width=0.23\textwidth]{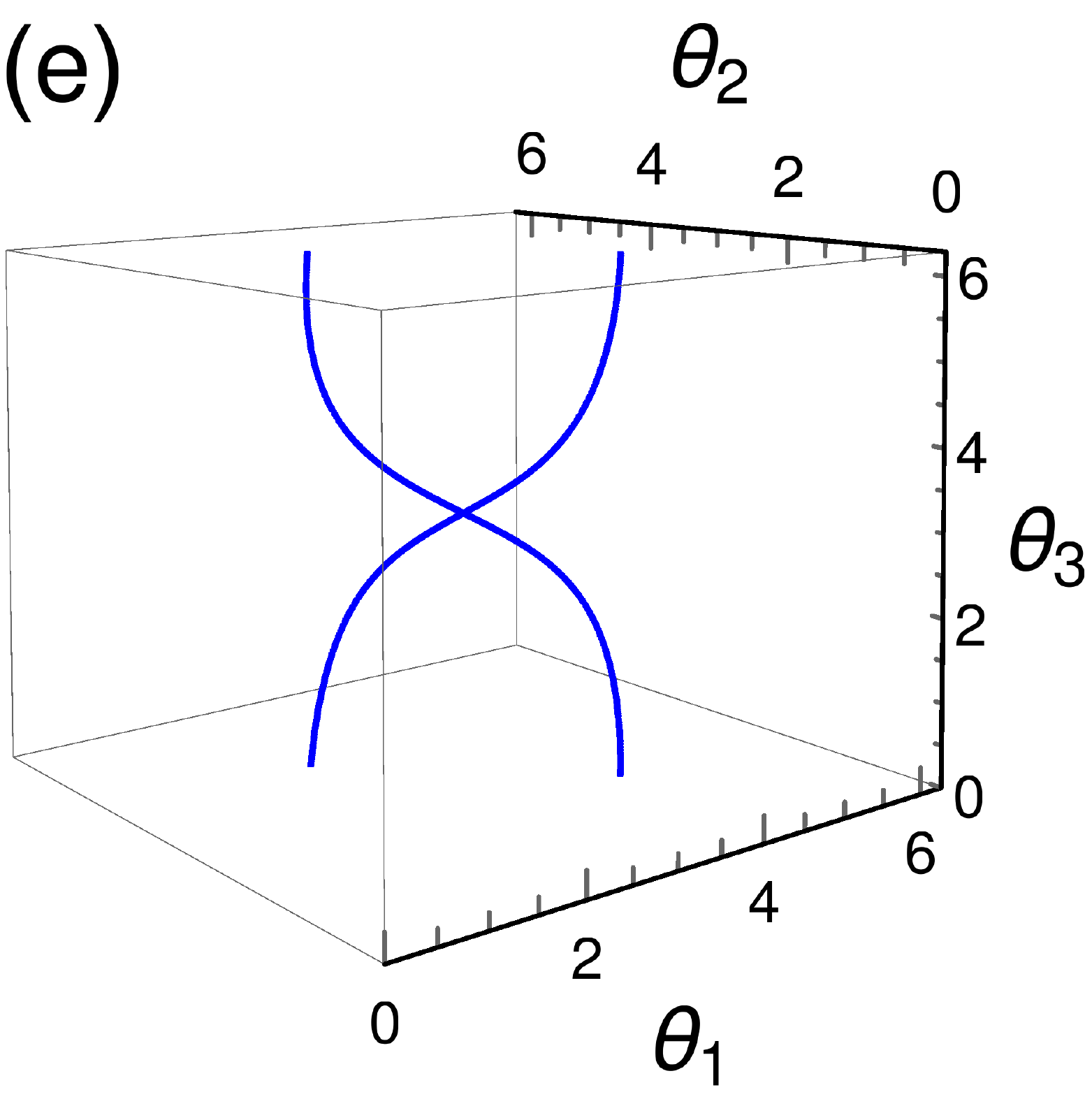}
\includegraphics[width=0.23\textwidth]{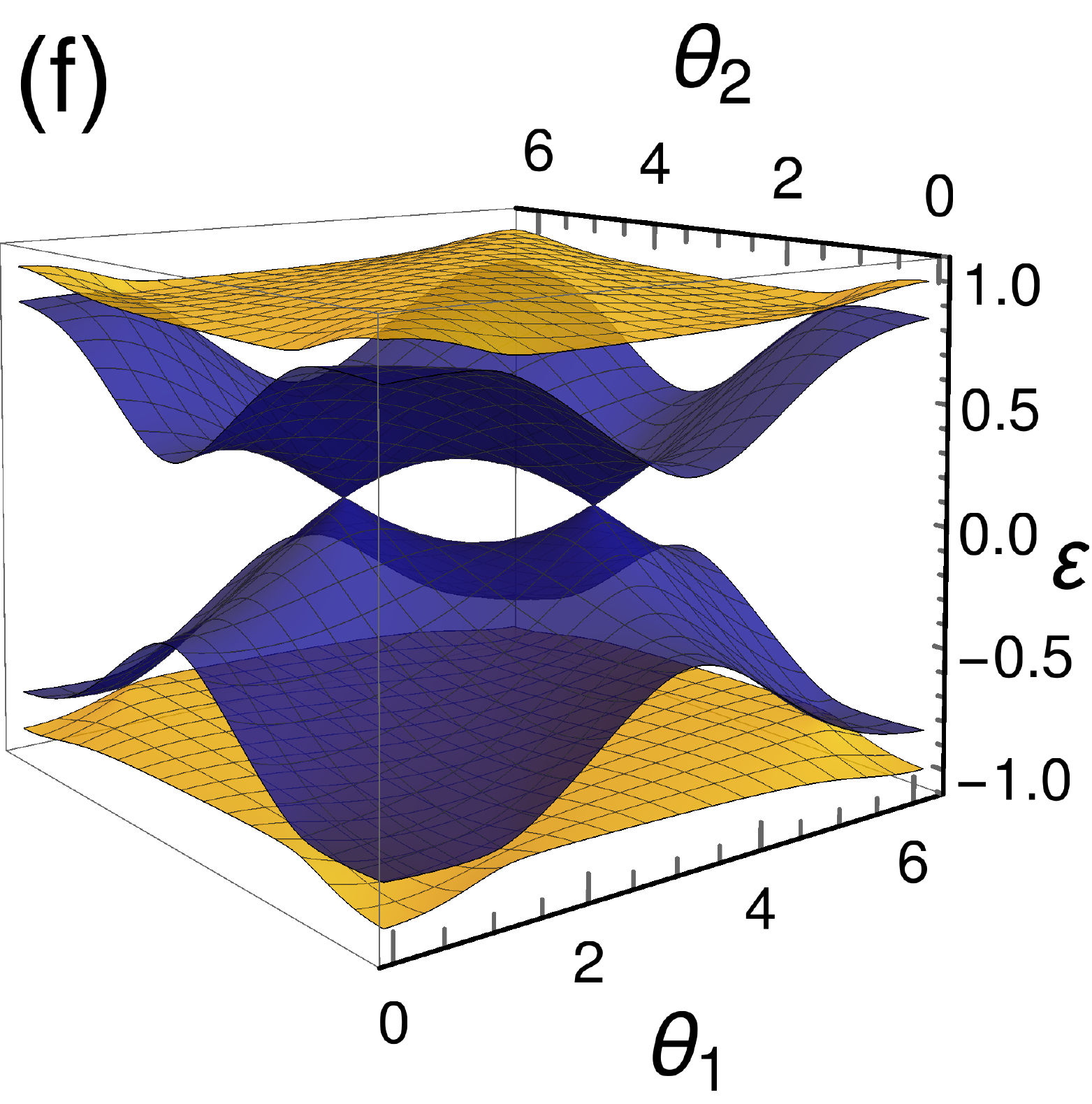}
\caption{[Color online] Energy spectrum of Andreev bound states for a four-terminal Josephson junctions [e.g. Fig. \ref{Fig-setup}]. In panels (a)-(d) we take the normal-region scattering matrix parameters [Eq.~\eqref{S}] $a=1/4$, $b=1/\sqrt{3}$, $c=1/\sqrt{2}$, $d=4/5$, $\varphi_1 = \pi/6$, and $\varphi_2 = \pi/3$. 
(a) Chern number of Andreev bands as a function of $\theta_3$ as computed from Eqs. \eqref{ABS} and \eqref{C}.  (b) An example of topologically trivial ($C_{12}=0$) gapped spectrum at $\theta_3 = \pi/3$. (c) An example of topologically nontrivial ($C_{12}=1$) gapped spectrum at $\theta_3 = 3\pi/4$. (d) An example of a nodal spectrum for $\theta_3 \approx 1.798$. In panels (e) and (f) we take $a=\varphi_1=\varphi_2=0$, $b=1/\sqrt{3}$, $c=1/\sqrt{2}$, and $d = 1$. (e) The trajectory of the nodal points in $\boldsymbol{\theta}$ space. (f) An example of a gapless spectrum at $\theta_3 = \pi/2$ exhibiting a pair of nodal points at $(\theta_\pm,\theta_\mp)$ with $\theta_\pm= 2\pi -\arccos{(-3/\sqrt{10})}\pm \arccos{(2/\sqrt{10})}$. } 
\label{Fig-ABS}
\end{figure} 

We find rich landscape of different band structures and determine that four-terminal devices can realize both Dirac and Weyl type singularities in ABS spectra. The band topologies can be characterized by the Chern number as a function of the phase $\theta_3$ computed according to the standard procedure by integrating Berry curvature over the unit cell spanned by phases $\theta_{1,2}\in[0,2\pi]$ 
\begin{align}\label{C}
& C_{12}(\theta_3) = \frac{1}{ 2 \pi} \iint_{0}^{2\pi} \! d\theta_{1} d\theta_{2} \, B_{12}(\theta_1,\theta_2;\theta_3), \nn \\
& B_{12}(\theta_1,\theta_2;\theta_3) =  -2 \sum_{k} \mathrm{Im} \langle  \partial_{\theta_{1}} \psi_k | \partial_{\theta_{2}} \psi_k \rangle.
\end{align} 
with $|\psi_k  \rangle$ being the corresponding bound state $k$. The complexity of emergent bands and variety of their shapes is dictated by a large number of parameters embedded in the scattering matrix. Thus we discuss several representative examples. 

\subsection{Weyl nodes}\label{Sec-Weyl}

Consider first the case with incommensurate choice of parameters $a=1/4$, $b=1/\sqrt{3}$, $c=1/\sqrt{2}$, $d=4/5$, and also two scattering matrix phases $\varphi_1=\pi/6$ and $\varphi_2=\pi/3$. The results for ABS bands and their Chern numbers as computed from Eqs. \eqref{ABS} and \eqref{C} are shown in Fig. \ref{Fig-ABS}(a-d).  As phase $\theta_3$ is tuned from zero the spectrum is overall gapped [Fig. \ref{Fig-ABS}(b)] and topologically trivial see Fig. \ref{Fig-ABS}(a). This regime persists until a certain critical value of $\theta_3\approx1.798$ is reached when Andreev bands touch at a single point as shown in Fig. \ref{Fig-ABS}(d), which is equivalent to a Dirac semimetallic phase. With the further increase of $\theta_3$ the gap reopens, see Fig. \ref{Fig-ABS}(c), but the system enters into the topologically nontrivial regime with $C_{12}=+1$. At a different point in a parameter space of phases bands would meet again closing and reopening the gap indicating another quantum phase transition to a state with different topological charge $C_{12}=-1$ as illustrated in Fig. \ref{Fig-ABS}(a). The corresponding band structures are not shown in the plot for brevity as they look alike.          

A different behavior is found in the case of more symmetric coupling when $a=0$, $b=1/\sqrt{3}$, $c=1/\sqrt{2}$, $d=\cos\chi$ and $\varphi_{1,2}=0$. This implies that $\alpha=0$ superconducting terminal couples to the other terminals equivalently and there is no reflection at this lead. This case realizes a pair of isolated Weyl singularities at $(\theta_1,\theta_2)=(\theta_\pm,\theta_\mp)$ with $\theta_\pm=2\pi-\arccos(-3/\sqrt{10})\pm\arccos(2/\sqrt{10})$ as shown in Fig. \ref{Fig-ABS}(f) for $\theta_3=\pi/2$ and $\chi=0$. For a different values of $\theta_3$ but still $\chi=0$ the location of the Weyl nodes can be computed explicitly and given by equations $3(\cos\theta_1+\cos\theta_2)+2\cos(\theta_1-\theta_2)+4=0$ and $1+3\cos\theta_3+2\frac{\cos(\theta_1-\theta_2)}{\cos^2[(\theta_1-\theta_2)/2]}=0$, for $\theta_1+\theta_2\in[2\pi,4\pi]$ when $\theta_2\in[0,\pi]$, and $\theta_1+\theta_2\in[0,2\pi]$ when $\theta_3\in[\pi,2\pi]$. These two lines of nodal points are plotted in Fig. \ref{Fig-ABS}(e). There is a very special high symmetry point of the model where these lines intersect, which corresponds to a single quadratic band touching between the Andreev bands.    

It is of importance to discuss experimental techniques that can probe properties of ABS directly or infer them indirectly from transport characteristics of a given device. Scanning tunneling and tunneling probe microscopies are the power imaging tools that can reveal complexities of the ABS spectra in various hybrid systems. Multiple experiments have been carried out recently on different heterostructures and proximity circuits involving superconducting and/or topological materials. Certainly, it will be fruitful to have such experiments done in multiterminal Josephson devices in a search for topological phases we described above. We hope that our results may stimulate such efforts. However, with the pure tunneling probe it may be challenging to identify transitions between different sectors of topologies. For that reason we will study transport probes as well. The most natural observables would be two-terminal Josephson current and conductances that in a multiterminal devices can be controlled and tuned by the other leads.   

\begin{figure}
\includegraphics[width=0.23\textwidth]{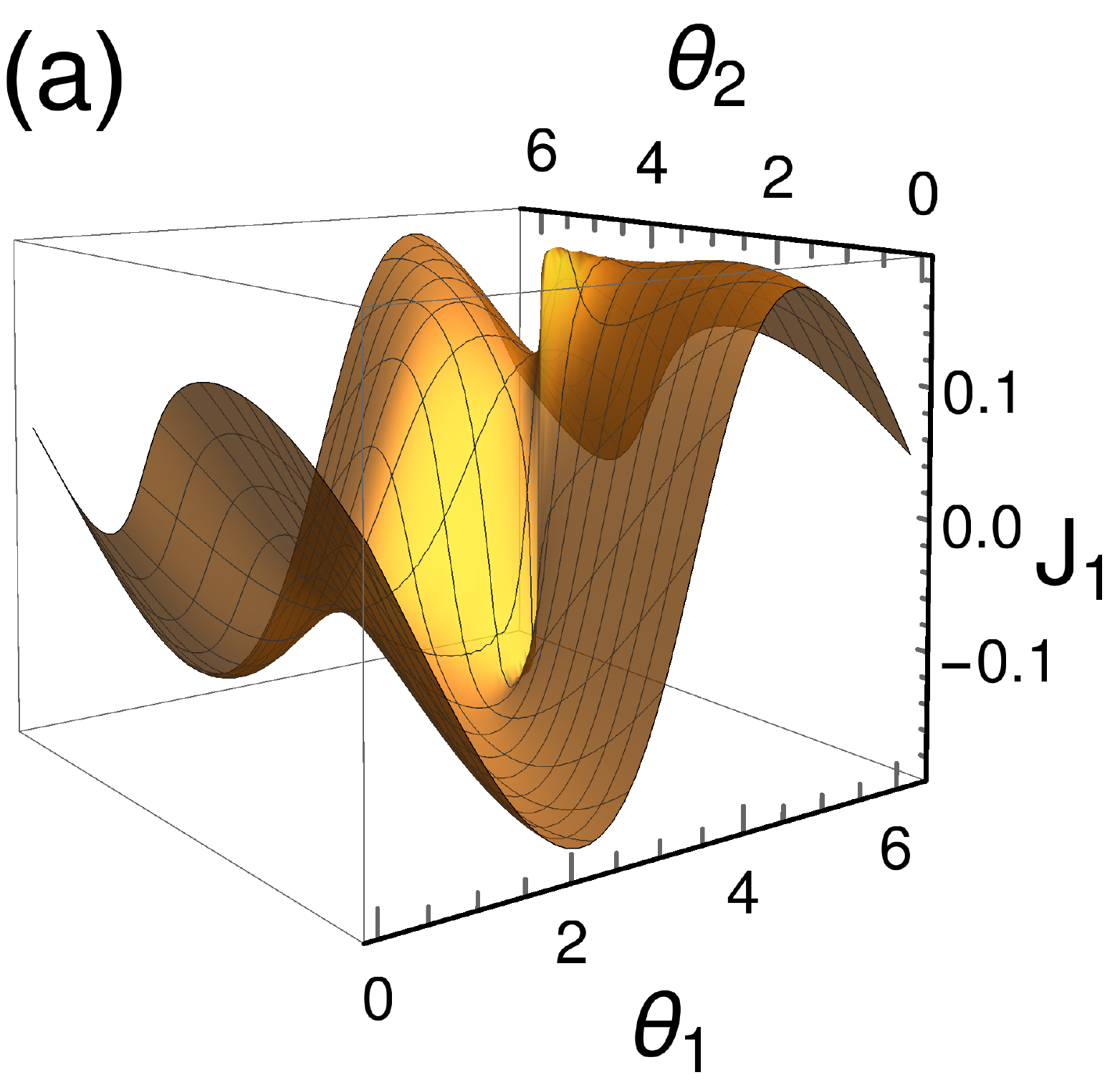} 
\includegraphics[width=0.23\textwidth]{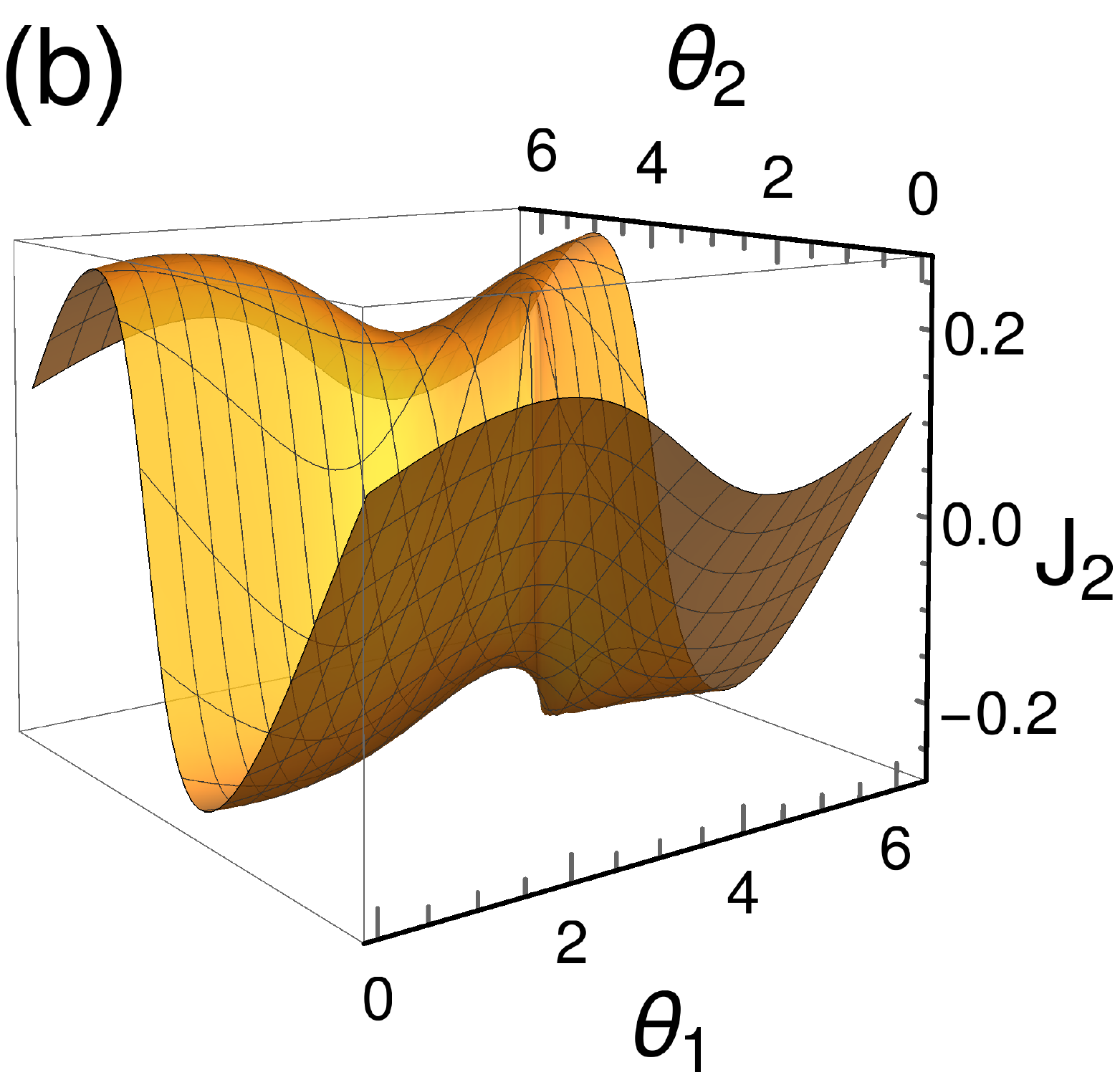} \\
\includegraphics[width=0.235\textwidth]{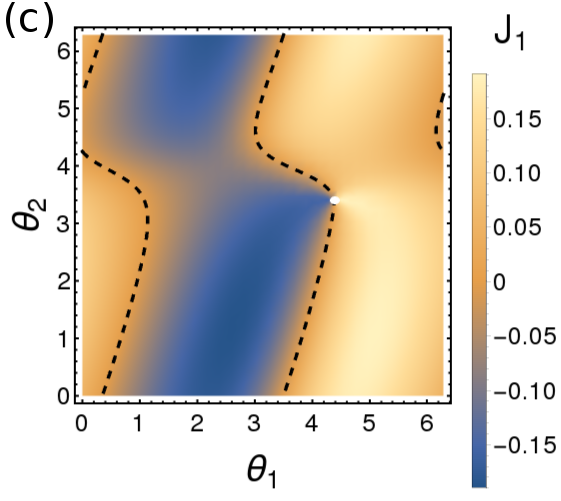} 
\includegraphics[width=0.235\textwidth]{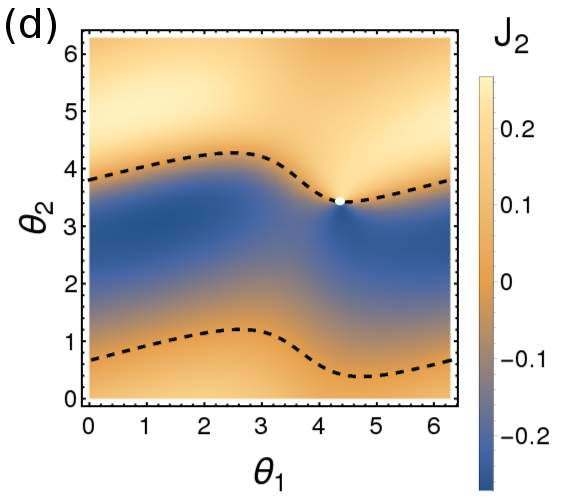} \\
\includegraphics[width=0.23\textwidth]{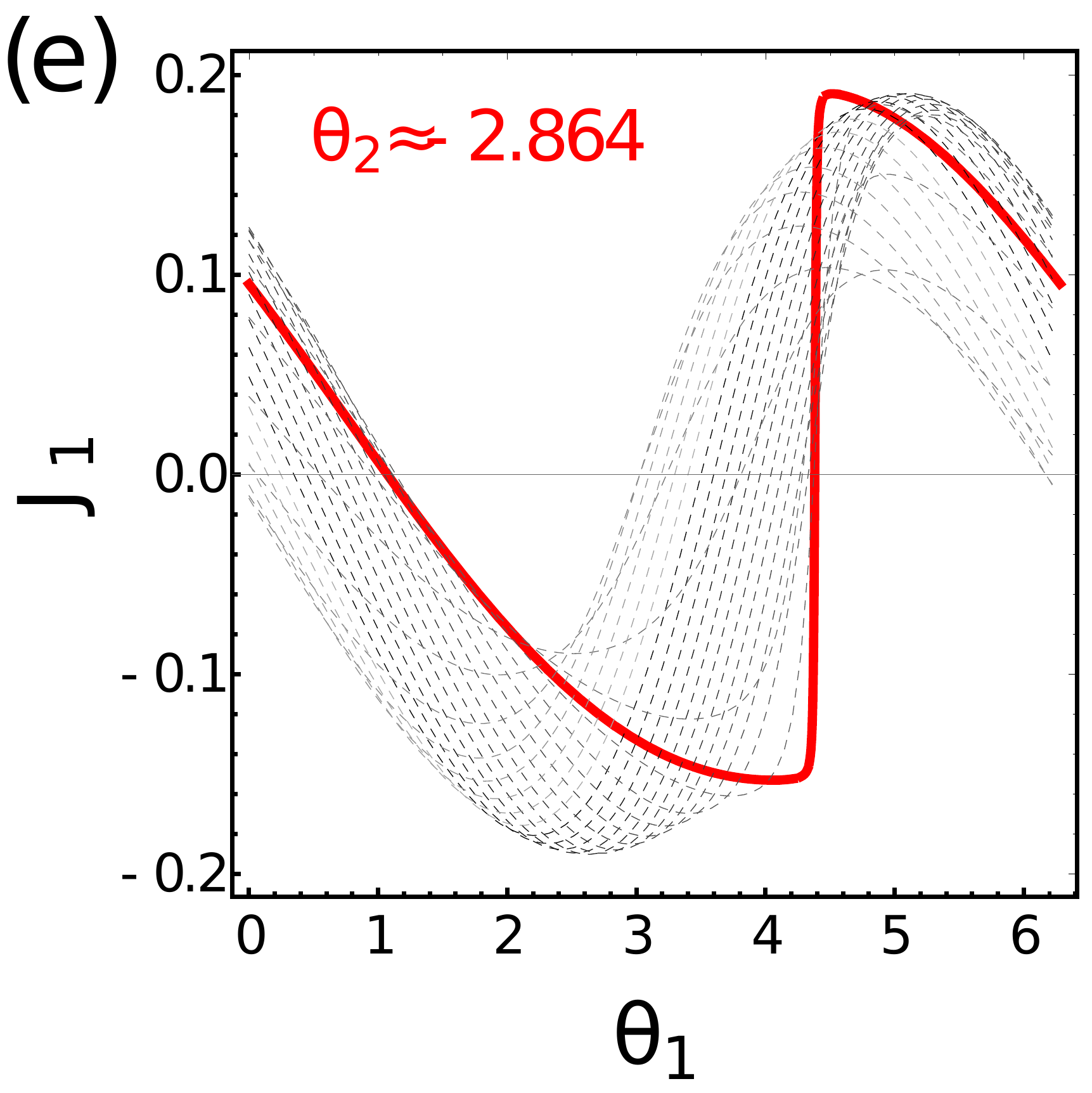} 
\includegraphics[width=0.23\textwidth]{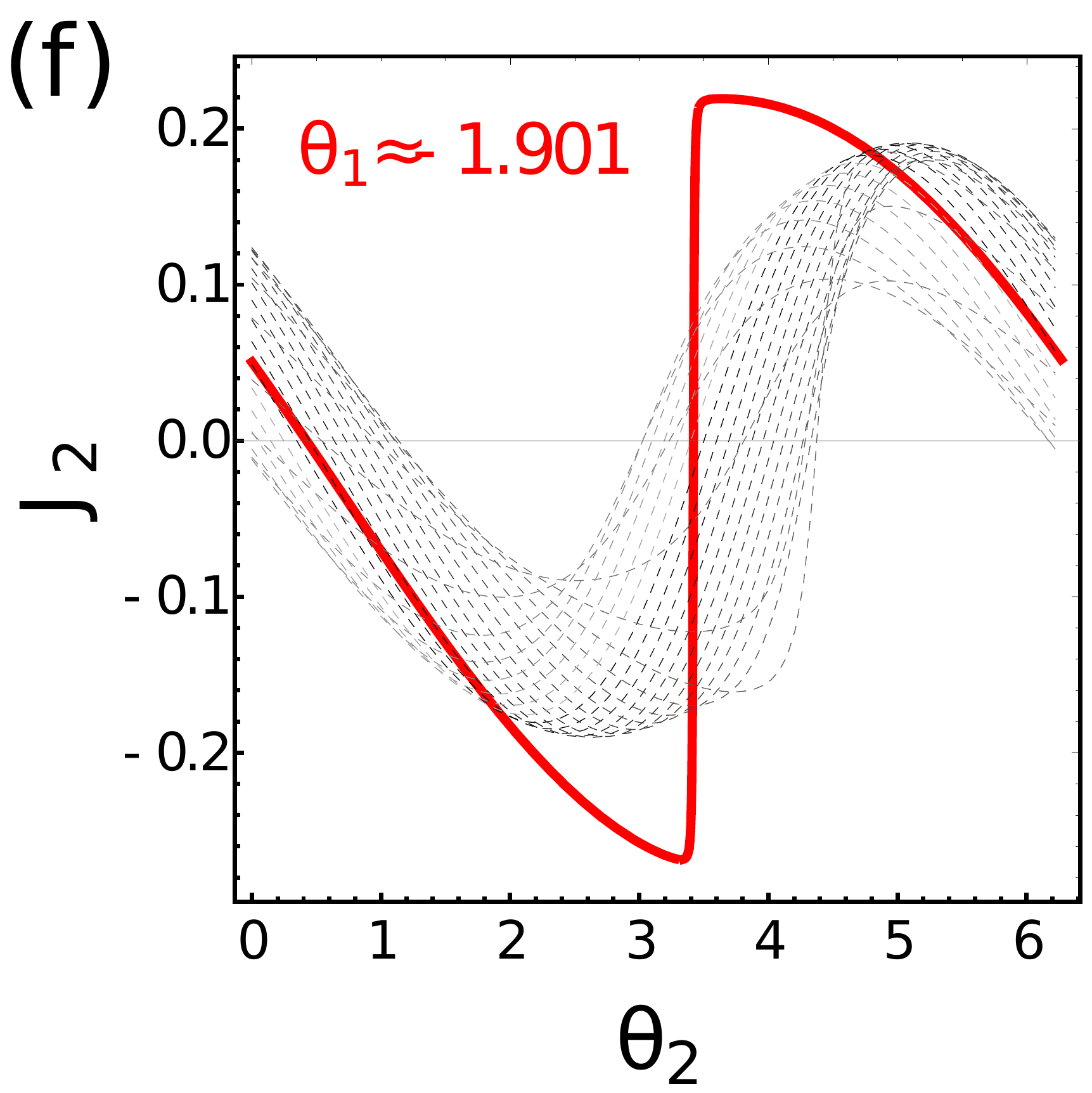} \\
\caption{[Color online] Josephson currents $J_{1,2}$ as functions of $\theta_{1,2}$ when $\theta_3$ tunes the system to the Weyl poin ($J_{1,2}$ are plotted in the unit of $2e\Delta/\hbar$). We take the same scattering matrix parameters as those in Fig.~\ref{Fig-ABS}(a,b) and fix $\theta_3=1.798$.  (a) and (b) Three-dimensional plots of $J_{1,2}(\theta_1,\theta_2)$. (c) and (d) Density plots corresponding to (a) and (b), respectively. Panels (e) [(f)] display $J_{1}$ ($J_{2}$) as a function of $\theta_1$ ($\theta_2$) for various values of $\theta_2$ ($\theta_1$). The red curves indicate traces of the Josephson currents cross the Weyl point.} 
\label{Fig-J1}
\end{figure} 

\section{Josephson supercurrents and conductance}\label{Sec-Jc-G}

In a scattering theory formalism the Josephson current through the lead $\alpha$ can be expressed terms of the Brouwer-Beenakker determinant formula [\onlinecite{BB}] (hereafter $k_B=1$)
 \begin{align}
J_\alpha &\, =  \frac{2eT}{\hbar} \frac{\partial}{\partial \theta_\alpha} \sum_{i\omega_n} \ln \mathrm{Det}\left[ \mathbb{I}_{n} - \gamma(\varepsilon) \, e^{i \hat{\theta}} \, \hat{s}^*(-\varepsilon) \, e^{-i \hat{\theta}} \, \hat{s}(\varepsilon) \right] \nn \\
     &\, = \frac{2eT}{\hbar}\sum_{i\omega_n} \frac{\partial_{\theta_\alpha} P_{n} [\gamma(i\omega_n);\hat{\theta}] }{ P_{n} [\gamma(i\omega_n);\hat{\theta}]},
\end{align}
where $\omega_n = (2 n +1) \pi T $ is the fermionic Mastubara frequencies, and we have used the polynomial representation for the determinant with the help of Eq.~\eqref{CUE-2n}. Formally factorizing the polynomial $P_{n}(\gamma) = \prod_{k=1}^{n/2}(\gamma-\gamma_k)(\gamma-\gamma_k^{-1})$ for $n \in \mathrm{even}$ and $P_{n}(\gamma) =(\gamma-1)P_{n-1}(\gamma)$ for $n \in \mathrm{odd}$, where $\gamma_k=e^{- 2 i \arccos{\varepsilon_k}}$ with $k$ labeling the Andreev bands, we obtain
\begin{align}
& J_\alpha = -\frac{2e}{\hbar} \sum_{k} \frac{\partial \varepsilon_k(\hat{\theta}) }{\partial \theta_\alpha}T\sum_{i\omega_n} g(i\omega_n), \nn \\
& g(i\omega_n) = \frac{8  \gamma(i \omega_n) \varepsilon_k(\hat{\theta}) }{\left[ \gamma(i\omega_n)-\gamma_k(\hat{\theta}) \right]\left[ \gamma(i\omega_n)-\gamma_k^{-1}(\hat{\theta}) \right]}.
\end{align}
Performing the Mastubara frequency summation via standard formula $T\sum_{i\omega_n} g(i\omega_n) = \sum_{z_0 = \pm \varepsilon_k} n_F(z_0)  \res[g(z_0)]$ where $n_F(z) \equiv (e^{\beta z}+1)^{-1}$ is the Fermi-Dirac distribution function with residues $\res[g(\pm \varepsilon_k)] = \pm1$, we obtain
\begin{align}\label{J}
J_\alpha = \frac{2e\Delta}{\hbar}  \sum_{k} \frac{\partial \varepsilon_k(\hat{\theta}) }{\partial \theta_\alpha}   \tanh\left[ \frac{\Delta}{2T}\varepsilon_k(\hat{\theta}) \right].
\end{align}  
This generic formula enables us to model supercurrents in conjunction with the shapes and topologies of the Andreev bands. This connection is most transparent in the zero temperature limit when $\tanh[\Delta\varepsilon_k(\hat{\theta})/2T]\to1$. 

\begin{figure}
\includegraphics[width=0.23\textwidth]{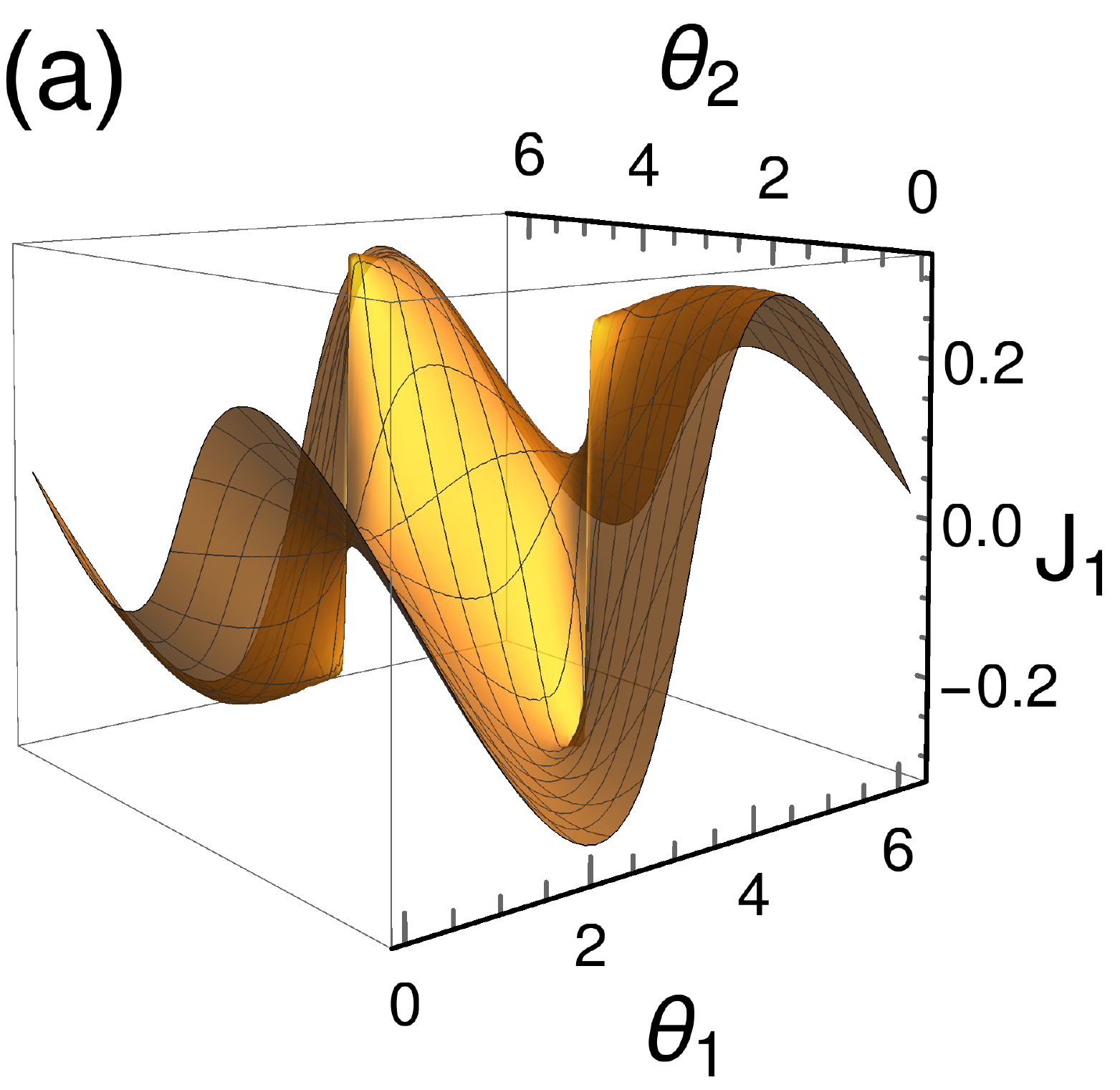} 
\includegraphics[width=0.23\textwidth]{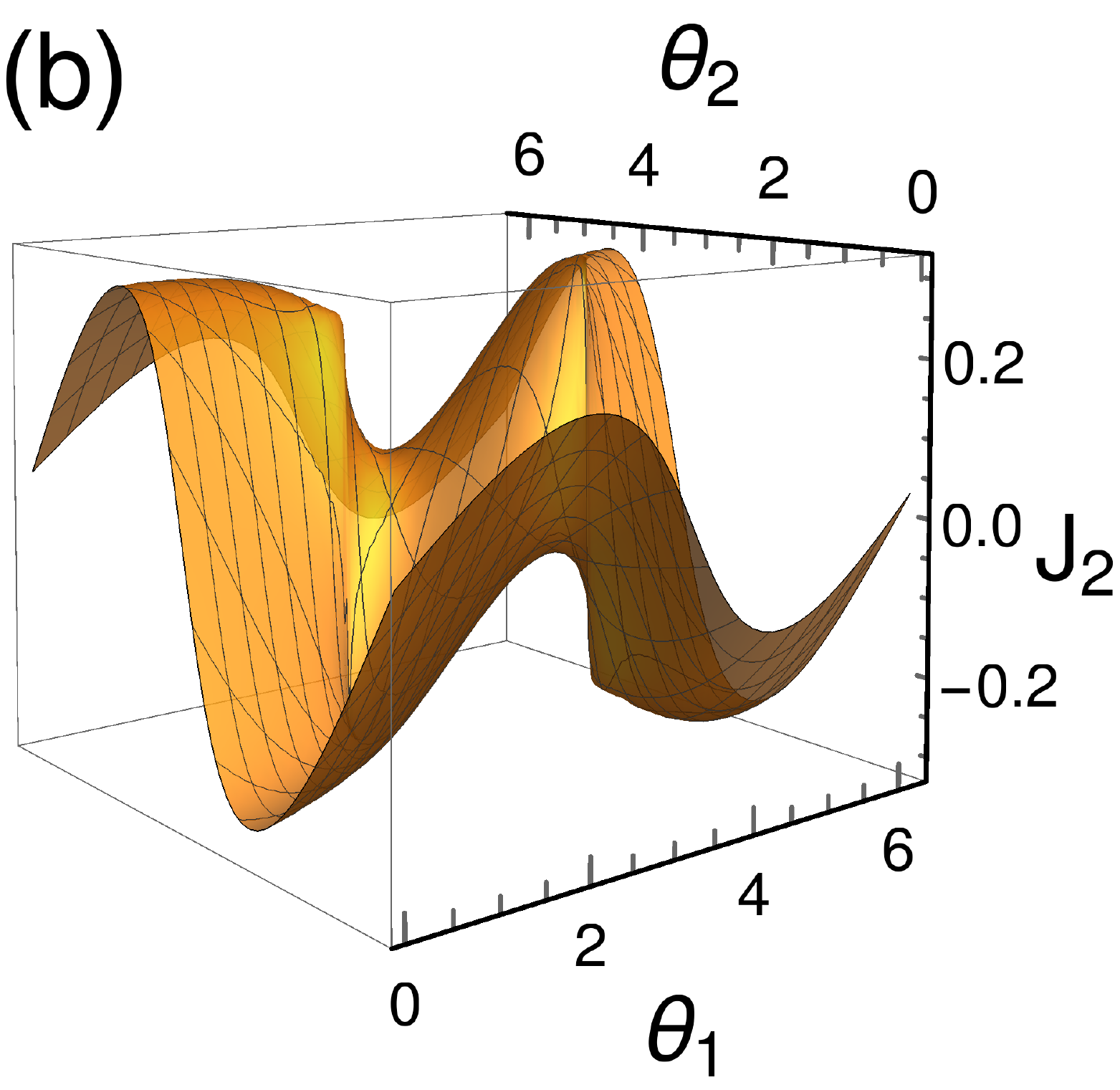} \\
\includegraphics[width=0.23\textwidth]{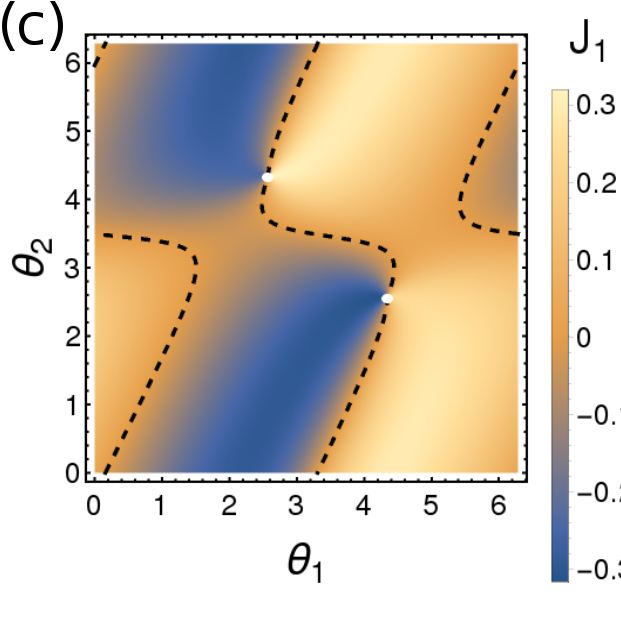} 
\includegraphics[width=0.23\textwidth]{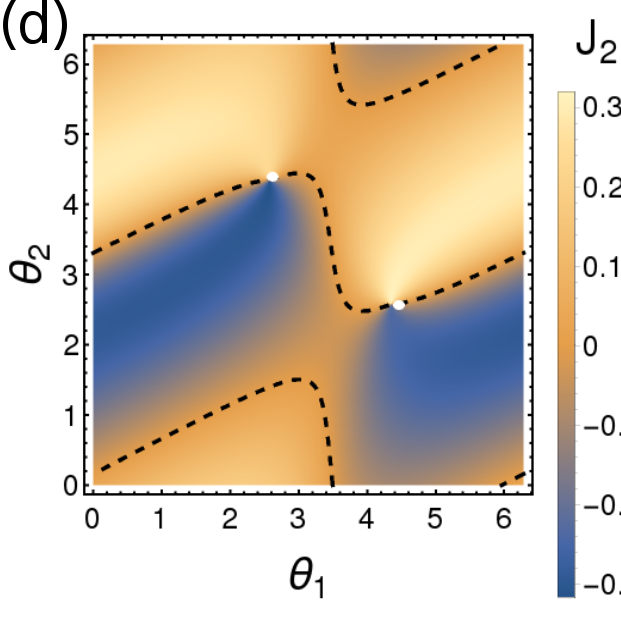} \\
\includegraphics[width=0.23\textwidth]{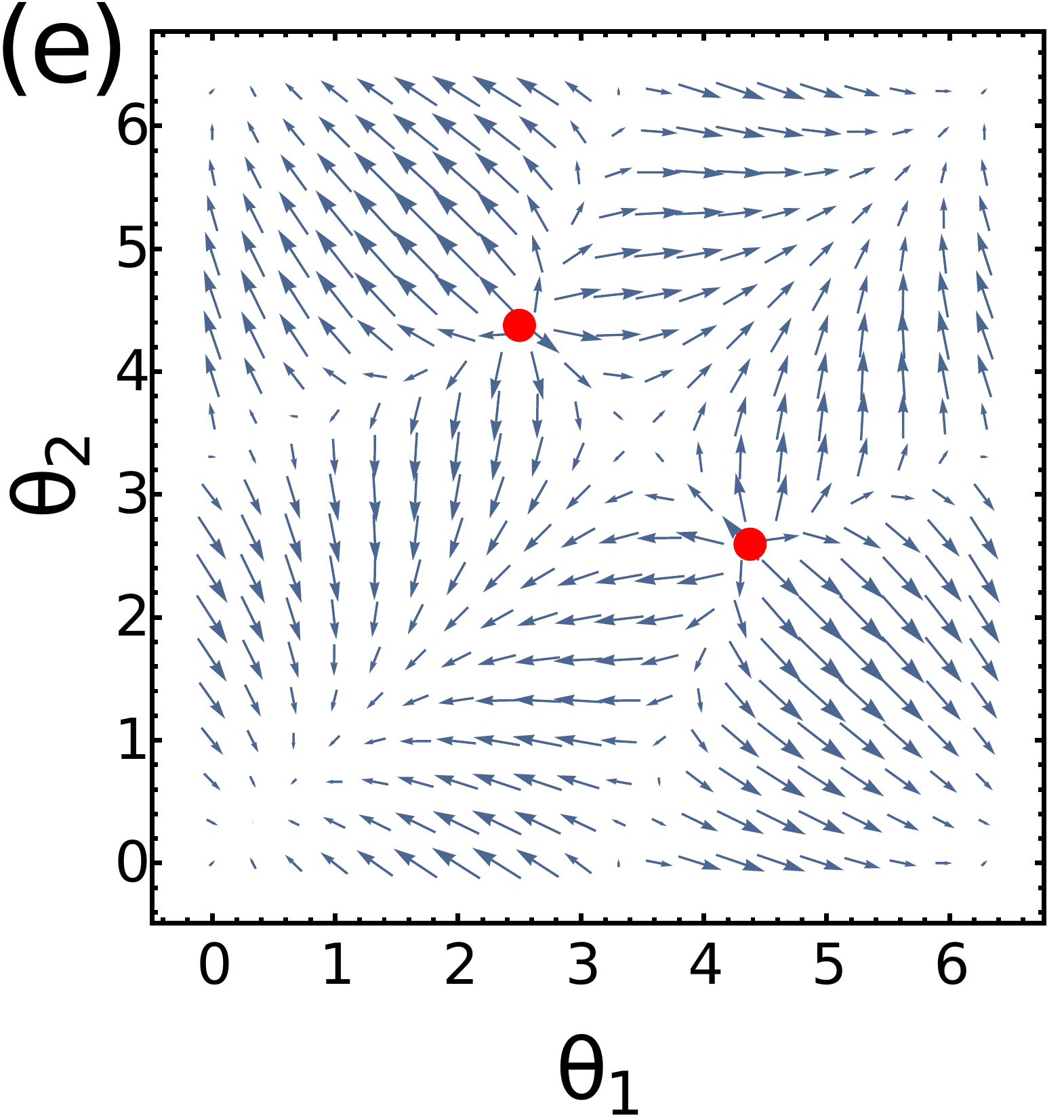} 
\includegraphics[width=0.23\textwidth]{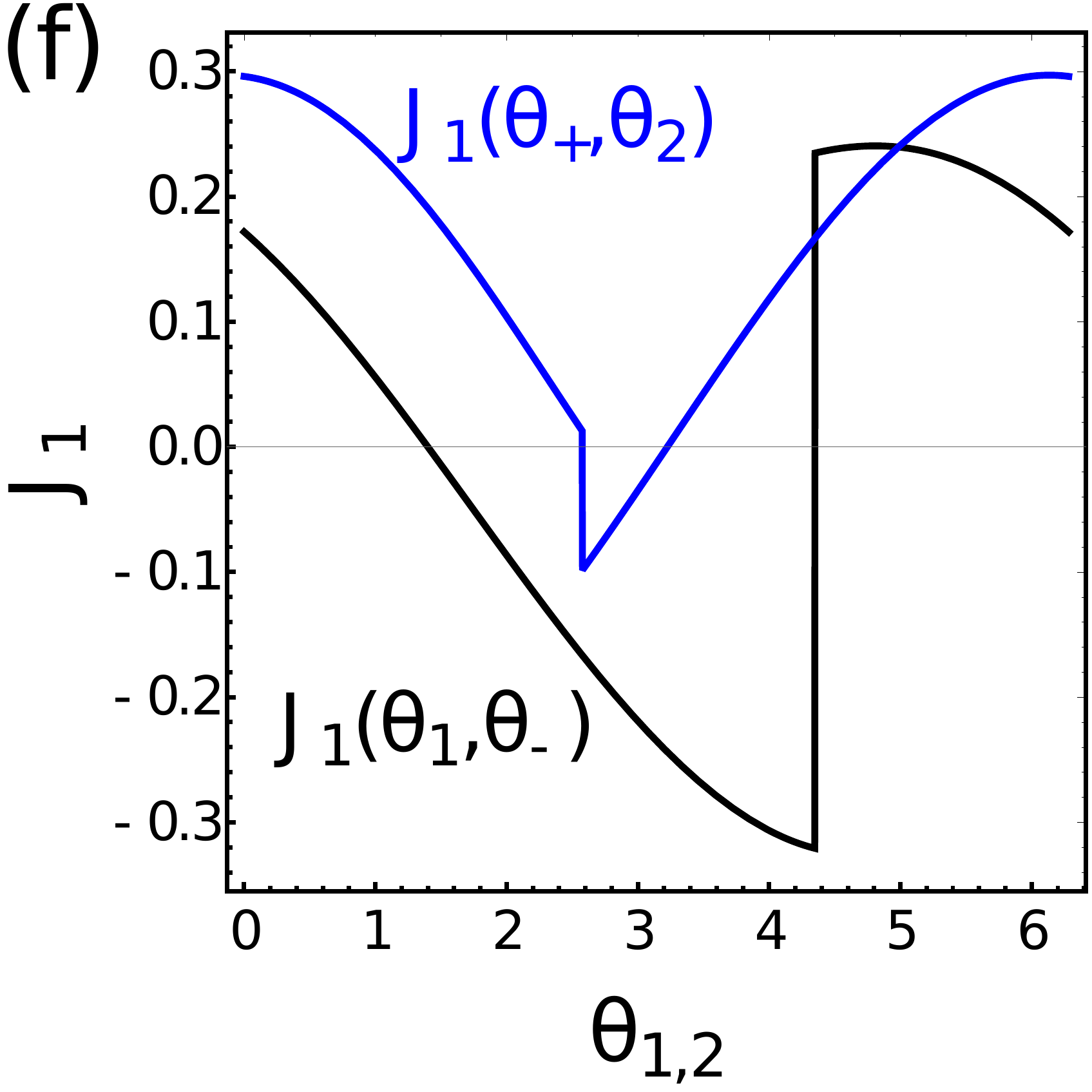} \\
\caption{[Color online] Josephson currents $J_{1,2}$ as functions of $\theta_{1,2}$ when the system exhibits a pair of nodal points. We take the same scattering matrix parameters as those in Fig.~\ref{Fig-ABS}(e) and (f). Panel (a) represent two-dimensional plot of the Josephson current in the first terminal as a function both phases, and panel (c) gives the same information but plotted differently to stress the sign change of the current and clearly identify location of the nodal points. Panels (b) and (d) give similar results for the current in the second terminal. Panel (e) shows votrex-like pattern of the current flow between the Weyl points. (f) Josephson current traces when device is tuned to one of the Weyl nodes $\theta_\pm$ and then one-dimensional cuts are made along either $\theta_1$ or $\theta_2$ directions across the other nodal point. Once singular point is passed current experience a discontinuous jump. } 
\label{Fig-J2}
\end{figure}

Figures \ref{Fig-J1}(a,b) display Josephson currents $J_{1,2}$ in two terminals as the system is tuned from topologically trivial gapped state to nodal gapless state by varying $\theta_3$. Plots on Fig. \ref{Fig-J1}(c,d) represent the same data but plotted in a different way for clarity to reveal regions of positive and negative currents as well as location of the nodal point. The series of one-dimensional cuts either in $\theta_1$ variable or $\theta_2$ show in Fig. \ref{Fig-J1}(e,f) how Josephson current changes as one tunes towards the singular point. Precisely at the node current exhibits discontinuous jump. We present similar results in Fig. \ref{Fig-J2} for the band structure of Fig. \ref{Fig-ABS}(f) with the pair of Weyl nodes. Fig. \ref{Fig-J2}(f) demonstrates similar discontinuities at both nodes $\theta_\pm$ and as a function of either $\theta_1$ or $\theta_2$ depending how the point is approached. A natural question arises of whether such nonanalytic behavior is exclusive. First we notice that as is clear from presented results the height of the jumps in Josephson CPR is not universal. It is also known that for certain types of Josephson junctions the current-phase relationship may be discontinuous [\onlinecite{CPR-JJ}]. For example, this happens in a long ballistic junctions. Other nonanalyticities in a form of the cusps or discontinuous derivatives of the current-phase relationship may survive even in disordered junctions or be specific to unconventional pairing states of superconductors forming the junction. In all these cases however nonanalyticities of CPR are pinned to a very specific values of phases. In contrast, here their location is variable and can be controlled experimentally by adjusting phases in the leads. 

What happens to be a distinguishing feature of a topological phase in multiterminal devices is a quantization of conductance in the adiabatic regime of ac Josephson effect. The quantization stems from the Berry curvature-induced correction to the phase velocity entering the current. To see this effect clearly consider deep subgap dc-voltages applied to superconducting terminals $eV_\alpha\ll\Delta$. In accordance with the second Josephson relation superconducting phases change with time as $\theta_\alpha(t)=2eV_\alpha t/\hbar$. To calculate the expectation value of the current operator $\hat{I}_\alpha=(2e/\hbar)\partial_{\theta_\alpha}\hat{H}$ in lead $\alpha$ it is convenient to introduce the basis of instantaneous wave functions of the time-dependent Bogolubov-de Gennes Hamiltonian $\hat{H}(t)$ such that $\hat{H}(t)|\psi_k(t)\rangle=\Delta\varepsilon_k(t)|\psi_k(t)\rangle$. By solving then the time-dependent evolution equation 
$i\hbar \partial_t|\psi_k(t)\rangle=\hat{H}(t)|\psi_k(t)\rangle$ to the first order in $\partial_t\theta_\alpha$ one finds time dependent current $I_\alpha(t)=I[\theta_\alpha(t)]$ in the form [\onlinecite{Riwar}]
\begin{equation}\label{I}
I_\alpha(t)\approx J_\alpha-\frac{4e^2}{\hbar}B_{\alpha\beta}V_\beta
\end{equation}
where the first term corresponds to Eq. \eqref{J} at $T=0$, whereas the second term is a correction proportional to the Berry curvature $B_{\alpha\beta}$ defined by Andreev band Bloch states in Eq. \eqref{C}. One should notice a similarity of this term to the anomalous velocity term of the current in the context if anomalous Hall effect. The time averaged current is equivalent to average over the Brillouin zone of ABS as over time phases $\theta_\alpha(t)$ uniformly sweep the unit cell $\theta_{1,2}\in[0,2\pi]$. In this case, the first term in Eq. \eqref{I} averages to zero as being pure gradient [Eq.~\eqref{J}], while the second term gets replaced by the Chern number in accordance with Eq.~\eqref{C}. As a result, one finds quantized conductance 
\begin{equation}\label{G}
\bar{I}_\alpha=G_{\alpha\beta}V_\beta,\quad G_{\alpha\beta}=-\frac{4e^2}{h}C_{\alpha\beta}. 
\end{equation}     
Stringent constraints have to be met for observability of this result as time dependence of superconducting phases will mediate non-adiabatic Landau-Zener transitions between Andreev bands giving rise to quasiparticle generation across the gap and corresponding dissipation [\onlinecite{Averin,Bardas}]. The severe constrain on required voltages to observe robust quantization comes from the fact that the probability of Landau-Zener tunneling $P_{LZ}$ is exponentially sensitive to voltages, namely $P_{LZ}=\exp[-\pi E^2_g/W]$, where $E_g$ is the gap of avoided crossing between Andreev states and $W$ measures the rate at which they approach each other, typically $W\sim eV\Delta$. These details were carefully analyzed in Ref.~[\onlinecite{Eriksson}].     

\section{Summary and outlook}\label{Sec-SumDis}

We have studied band topology of Andreev levels in multiterminal Josephson junctions focusing on a four-terminal devices. Depending on the properties of the normal region described by a scattering matrix connecting different terminals, Andreev bands reveal either Weyl or Dirac type nodal points in the parameter space of superconducting phases. In the limit of energy-independent scattering matrix, relevant for short weak links, the energy spectrum can be found exactly and given by Eq.~\eqref{ABS}. We find multiple quantum phase transitions between topological and trivial gapped states that can be quantified by Chern numbers (Fig.~\ref{Fig-ABS}). These transitions can be controlled and manipulated in experiments with Andreev interferometers [\onlinecite{Giazotto-1, Giazotto-2}]. Apart from the band structure that can be probed by tunneling spectroscopies, we also calculate transport characteristics including Josephson current-phase relationships [Eq.~\eqref{J}] and two-terminal conductance [Eq.~\eqref{G}].  

Multiterminal proximity circuits offer numersou opportunities for future research directions including intriguing connections to the high-energy physics. In the context of solid state systems, it is of interest to explore possibilities to mimic transport and perhaps also optical properties of Weyl semimetals in the regime of the ac Josephson effect. In particular, it has been shown recently that superconductivity provides an access to the chiral magnetic effects of an unpaired Weyl cone [\onlinecite{OBrien}]. When topological superconductor is introduced as an element of the multiterminal junction, such devices host Majorana states. One may explore perspectives of braiding them by winding superconducting phases within the unit cell of ABS with voltage pulses. Further applications to Josephson qubits are possible especially when normal region of the junctions is formed by a semiconducting material which enables efficient gating. Such hybrid semiconductor-superconductor gatemon-type qubits may allow an enhanced degree of gate control and performance [\onlinecite{DiCarlo,Marcus-1,Marcus-2}]. Finally, generalizations of the Dirac theory to higher-dimensional spaces described by second Chern class topological invariants of the corresponding SU$_2$($n$) group non-Abelian gauge fields proposed in high-energy theories [\onlinecite{Yang}] can be searched and perhaps realized with $n$-terminal Josephson devices.         

\section{Acknowledgments}

This work was financially supported by NSF Grant No. DMR-1606517 and in part BSF Grant No. 2014107 (H.X.), by NSF Grant No. DMR-1653661 and Vilas Life Cycle Professorship program (A.L.), by ARO Grant No. W911NF-15-1-0248 (M.V.), and the Wisconsin Alumni Research Foundation.

\end{document}